\PassOptionsToPackage{sort&compress}{natbib}
\documentclass{elsarticle}
\usepackage{float}
\usepackage{graphicx} 
\usepackage{subcaption}
\usepackage{gensymb}
\usepackage{amsmath}
\usepackage{xcolor}
\usepackage{longtable}
\usepackage{url}
\usepackage[export]{adjustbox}
\usepackage{hyperref}

\begin{document}
\begin{frontmatter}
\title{A new restitution model for wet particle collisions using fully resolved numerical simulations}
\author[2,1]{A.H. Huijgen}
\author[3]{F. Bunke}
\author[3]{S. Heinrich} 
\author[4,1]{Y. Tang}
\author[4,1]{N.G. Deen}
\author[2,1]{M.W. Baltussen\corref{cor1}}
\address[2]{Multiphase Flows for Energy Applications, Department of Chemical Engineering and Chemistry, Eindhoven University
of Technology, P.O. Box 513, 5600 MB Eindhoven, The Netherlands \fnref{ 2}}
\address[3]{Institute of Solids Process Engineering and Particle Technology, Hamburg University of Technology, Denickestraße 15, Hamburg, 21073, Germany\fnref{ 3}}
\address[4]{Power and Flow, Department of Mechanical Engineering, Eindhoven University
of Technology, P.O. Box 513, 5600 MB Eindhoven, The Netherlands \fnref{ 4}}
\address[1]{Eindhoven Institute for Renewable Energy Systems, Eindhoven University of Technology, Eindhoven, The Netherlands, P.O. Box 513, 5600 MB Eindhoven, The Netherlands \fnref{ 1}}

\ead{m.w.baltussen@tue.nl}
\cortext[cor1]{Corresponding author}\date{November 2025}

\begin{abstract}
Particle collisions play a central role in wet particle processes such as granulation, drying, and chemical reactors, yet the influence of the liquid on the collision dynamics remains insufficiently understood. 
This study investigates the influence of liquid viscosity, liquid density, liquid content, impact velocity, solid density, and surface tension on the wet restitution coefficient using numerical simulations. 
In this study, a Direct Numerical Simulation (DNS) framework is used that has been validated against dedicated experiments of wet-wet particle collisions, showing a good agreement for both liquid-bridge evolution and rebound dynamics. 
The restitution coefficient decreases with increasing liquid viscosity, liquid density, and liquid-to-solid density ratio, whereas increasing impact velocity increases rebound. 
The negligible influence of surface tension indicates that viscous dissipation is the dominant source of energy loss under the investigated conditions.
As existing Stokes number formulations were found to only capture part of the governing physics, a generalized Stokes number is proposed to incorporate both the viscous and inertial response time scales.
A modified restitution coefficient model based on the generalized Stokes number successfully collapses the numerical and experimental data onto a single trend and predicts the wet restitution coefficient with an RMSE of only 4.99\%.
\end{abstract}

\end{frontmatter}
\section{Introduction}
The fluidization of granular materials is widely used in chemical processes, biotechnology, granulation, and pharmaceuticals \cite{FAN2024119308, SHAO2022117090}. 
The enhanced mixing in such systems is primarily driven by frequent particle-particle collisions. 
Although the particles are primarily suspended in a gas flow, liquid can also be present as a liquid binder, reagent, or as a residue of drying. 
The presence of liquid drastically alters the collision dynamics, effectively increasing the energy dissipation. 
The loss of kinetic energy can ultimately lead to agglomeration of the particles in large complexes. 
These agglomerates are either desired, e.g. in granulation, or undesired, e.g. in drying and specific chemical reactions. 
Nevertheless, understanding the onset of agglomeration and the reduction in momentum, typically expressed via a restitution coefficient, is essential to understand the whole process.\\

In the literature, several model systems have been studied to quantify the effect of liquid addition on the reduction of the restitution coefficient.
One commonly employed model system is the Newton cradle, in which wet particles suspended as pendulums undergo guided collisions. 
For highly viscous liquids ($\mu_l$ = 1-10 $Pa\cdot s$), the restitution coefficient exhibits a clear dependence on the Stokes number, but the exact relation differs depending on the input conditions \cite{DANCZYK2022117112,Donahue2010, donahue2010a,Donahue2012,  Punch2023, SHAO2022117090}. 
\citet{ROYCHOWDHURY2026121816} used an air table to investigate the collision between wet particles and similarly observe a decrease in restitution coefficient with decreasing Stokes number (i.e. with decreasing impact velocity and viscosity) and increasing liquid content. 
Although these experiments show the interactions between particles, the collisions are guided, which is not representative for the particle interactions during fluidization. 
Another commonly studied model system is particle-plate collisions, in which a dry particle impacts on a plate covered by a thin liquid layer\cite{Antonyuk2009InfluenceImpact,  Buck2019CollisionExperiments, CRUGER2016209, DAVIS_RAGER_GOOD_2002, gollwitzer, Ma2013StokesNumber, Ma2016} or a wet particle collides with a dry plate \cite{Buck201821}.
Similarly to Newton cradle experiments, these studies showed that the restitution coefficient decreases with the Stokes number for process fluids such as water and glycerol. 
To account for the liquid layer thickness, later studies incorporated this parameter directly into the Stokes number formulation \cite{Ma2013StokesNumber, Ma2016, Buck2019CollisionExperiments}. 
These systems have also been investigated using numerical simulations \cite{FAN2025121253, Jain2012DirectMethod, SHAO2022117090, singh2026numericalstudycoefficientrestitution, SONG2025170823,  Tang2017}. 
 \citet{BUCK2017218} and \citet{CRUGER2016209} showed that the liquid content has a limited influence on the tangential restitution coefficient, whereas the liquid layer primarily affects the normal component.\\

Although these systems provide valuable insight into normal restitution coefficients, they do not fully capture the dynamics of collisions between two freely moving particles because of experimental challenges related to wetting, controlled release, and rotational measurement.
Only a limited number of studies have experimentally investigated collisions between freely moving particles \cite{Bunke2024a, Chandramohan2005, Foerster1994, Labous1997, Lorenz1997}, mainly focusing on the development of reliable experimental techniques. 
\citet{Bunke2024a} showed that the restitution coefficient is significantly lower for wet particles than for dry particles and decreases with increasing liquid content.\\

Numerically studies have investigated wet particle collisions. 
Collisions between a partially wetted particle and a dry particle exhibit distinctly different behavior compared to wetted particle–wall collisions, as reported by \citet{SHAO2022117090}, highlighting the need to investigate these systems separately. 
The same study further shows that the restitution coefficient decreases with increasing liquid content and viscosity.
This is confirmed for the collision between a partially wetted particle and a partially wetted wall, where a reduction in rebound velocity results in an increase in liquid viscosity \cite{PAN2026123656}.
In \citet{Kan2015, Kan2016}, it was further shown that the restitution coefficient increases with increasing impact velocity. 
Similar trends are observed for collisions between a dry and a fully covered particle, where the restitution coefficient increases with increasing Stokes number \cite{FAN2024119308}. 
In the extension by \citet{FAN2025121253}, where both particles are covered with liquid, it is shown that the restitution coefficient decreases with decreasing Weber number, Reynolds number, and layer thickness, resulting in a well-defined agglomeration boundary within a three-dimensional parameter space. 
This behavior can be attributed to increasing in pressure and viscous forces with increasing impact velocity, liquid film thickness and viscosity \cite{WANG2025120546}. 
\citet{huijgendrywetcollision} performs a close comparison with the experimental work of \citet{Bunke2024a} to validate the numerical framework. 
In addition, the results show that the viscous dissipation directly after the collision is the main source of energy dissipation.\\

Despite these efforts, no model has been developed to predict the outcome of the collision of freely moving wet particles based on initial conditions and all physical properties.
To develop a unifying model for the normal restitution coefficient, simulations are performed for wet particle collisions under varying conditions.
First, the numerical system is validated against new experiments for two wet particles and at lower Stokes number than those reported by \citet{Bunke2024a}. Using the validated numerical framework, viscosity, solid and liquid density, impact velocity, and liquid layer are systematically varied. Subsequently, the simulations are unified using a modified Stokes number that incorporates both viscous and inertial effects.

\section{Numerical methods}
The numerical framework is based on  \citet{huijgendrywetcollision}. In this section, a brief overview is provided and the interested reader is referred to \citet{huijgendrywetcollision} for further details. 
The fluids involved in this three-phase system are assumed to be Newtonian, incompressible, immiscible, and isothermal. 
A one-fluid formulation is used where the fluid viscosity and density are averaged harmonic and arithmetic averaging based on the liquid fraction, respectively. 
The governing equations are the continuity and Navier Stokes Equations, given in Eqs.~\ref{paper.eq:continuity} and \ref{paper.eq:navierstokes}. 
A no-slip boundary condition is imposed at the solid surface, assuming a non-porous material. 
\begin{equation}
\nabla \cdot \boldsymbol{u} = 0 \label{paper.eq:continuity}
\end{equation}
\begin{equation}
    \rho\frac{\partial \boldsymbol{u}}{\partial t}  = -\nabla p - \rho\nabla \cdot (\boldsymbol{uu}) {-}  \nabla\cdot\boldsymbol{\tau} + \rho\boldsymbol{g}+\boldsymbol{F_{\sigma}} \label{paper.eq:navierstokes}
\end{equation}
Here, $\boldsymbol{u}$ and $p$ are the velocity and the pressure, the fluid stress tensor is defined as $\boldsymbol{\tau} = -\mu[\nabla \boldsymbol{u} + (\nabla \boldsymbol{u})^T]$, and $\boldsymbol{g}$ and $\boldsymbol{F_{\sigma}}$ are the gravitational constant and a body force representing the surface tension, respectively. 
These equations are solved using a projection-correction method. 
In the prediction step, the Navier-Stokes equations are solved using a explicit second-order MINMOD scheme for convection \cite{HARTEN1983357, VanLeer1977} and a semi-implicit second-order central difference scheme for the viscous stress term. The equations are decoupled following the decoupling of the stress tensor \citet{Dijkhuizen2010}. 
The system is preconditioned using an incomplete LU-decomposition preconditioner from the Ifpack2 package of the Trilinos project \cite{TeamTheTrilinosProject2023TheWebsite} and solved using the Biconjugate gradient stabilized method (BiCGSTAB2) \cite{VanderVorst1992}. To ensure that the velocity field adheres to Eq. \ref{paper.eq:continuity}, a Poisson equation for the pressure correction $\delta p = p^{n+1}-p^n$ is solved (Eq. \ref{eq:method:presurecorection}) in combination with Eq. \ref{eq:method:presurecorection:velocitypdate}. Here, subscript $p^{n}$ and $p^{n+1}$ denote the previous and the current time steps, respectively, and $\boldsymbol{u}^*$ is the predicted velocity. The Poisson equation is stiff and is solved using an algebraic multigrid preconditioner from the MueLu package of the Trilinos project \cite{TeamTheTrilinosProject2023TheWebsite}.
\begin{equation}
    \label{eq:method:presurecorection}
 \nabla \cdot \left[ \frac{\Delta t}{\rho^n}\nabla(\delta p) \right] = \nabla\cdot \boldsymbol{u}^*
\end{equation}  
\begin{equation}
    \label{eq:method:presurecorection:velocitypdate}
    \boldsymbol{u}^{n+1} = \boldsymbol{u}^* -  \frac{\Delta t}{\rho^n}\nabla \delta p
\end{equation}
In the simulations, two interface need to be represented.
The gas-liquid interface is represented with the Local Front Reconstruction Method (LFRM) \cite{Shin2011}. 
This front tracking method has been proven to outperform the Volume of Fluid method for binary droplet collisions at relatively course grid resolutions \cite{Huijgen2024a}. The interface is represented by a triangular mesh smaller than the Eulerian grid cells. These markers are advected using a 4th order Runga-Kutta scheme with velocity obtained from a cubic spline interpolation. To maintain mesh quality, the interface is regularly remeshed \cite{Shin2011} and smoothed \cite{Kuprat2001VolumeLines}. Coalescence and break-up are enabled by implementation of the tetra-marching procedure of \citet{Yoon2010}. The surface tension force $\boldsymbol{F_\sigma}$ in Eq.\ref{paper.eq:navierstokes} is calculated using the hybrid method of \citet{Shin2005AccurateMethod}. \\

The no-slip boundary at the fluid-solid interface is enforced using the a second-order ghost cell Immersed Boundary Method. Instead of employing a body-fitted mesh, that constantly needs to be re-meshed when the particles move, the no-slip boundary condition is directly imposed on the momentum equations. Ghost value is extrapolated using a second order polynomial and satisfies the imposed boundary condition on the solid surface. These are directly used in the discretization of the Navier-Stokes equations. Further details can be found in \citet{Deen2012DirectSystems}. 

The particle motion is governed by the drag force on the solid surface (Eqs. \ref{eq::TranslationBalanseModified} and \ref{eq::RotationalBalance}). 
\begin{equation}\label{eq::TranslationBalanseModified}
    m_p \frac{d \boldsymbol{v}_p}{dt} =- \oint_{\Gamma_m} (\boldsymbol{\tau \cdot n} + p \boldsymbol{n}) dS +  m_p \boldsymbol{g}
    \end{equation}
    \begin{equation}
 \label{eq::RotationalBalance}
    I_p \frac{d \boldsymbol{\omega}_p}{dt} = -\oint_{\Gamma_m} (\boldsymbol{r}_{\Gamma_m} - \boldsymbol{r}_p) \times (\boldsymbol{\tau \cdot n}  ) dS 
\end{equation}

\begin{figure}
    \centering
    \includegraphics[width=0.5\linewidth]{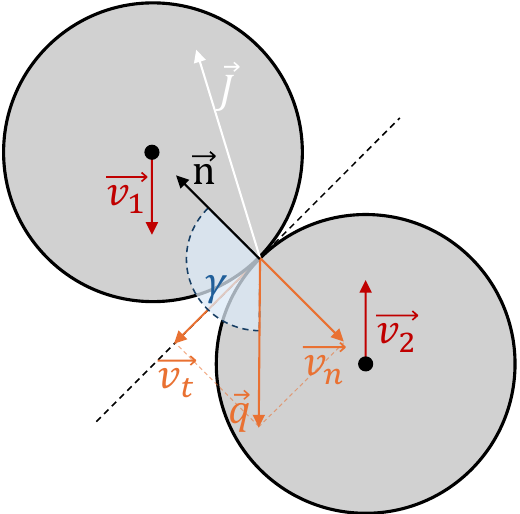}
    \caption{Schematic representation of hard-sphere collision.}
    \label{fig:collisionanglesketch}
\end{figure}

$m_p$ and $I_p$ are the mass and inertia of the object. The drag force consists of pressure forces ($ p \boldsymbol{n}$) and forces due viscous stresses ($\boldsymbol{\tau \cdot n}$), which can obtained by extrapolating the pressure and the velocity gradients to the solid surface \cite{Deen2014}, while $n$ is normal to the solid surface. The viscous drag uses the area correction from \citet{BubblesFlows}. The particle collisions are modeled using a hard-sphere approach \cite{Hoomans1996}. The change in the velocity and the associated slip velocity $q$ is defined as: 
\begin{align}
 \boldsymbol{v_i}' = \boldsymbol{v_i} \pm \frac{\boldsymbol{J}}{m_i} &&    \boldsymbol{\omega_i}' = \boldsymbol{\omega_i} - \frac{R_i}{I_i} (\boldsymbol{n} \times \boldsymbol{J})
\end{align} 
\begin{equation}
    \boldsymbol{q} = \boldsymbol{v_1} - \boldsymbol{v_2} - (R_1 \boldsymbol{\omega_1} + R_2 \boldsymbol{\omega_2}) \times \boldsymbol{n}
\end{equation} 
Where the prime symbol denotes the corresponding quantity after collision. The collision angle ($\gamma$) is the angle between the solid normal and the slip velocity $\boldsymbol{q}$, which is by definition 180\textdegree~for head-on and 90\textdegree~for fully oblique collisions, also illustrated in Figure \ref{fig:collisionanglesketch}.
The slip velocity before and after the collision ($\boldsymbol{q}, \boldsymbol{q'}$) and the exchanged momentum ($\boldsymbol{J}$) can be expressed using a normal restitution coefficient ($e$), tangential restitution coefficient ($\beta$), and an effective friction coefficient ($\mu$). 
\begin{equation}
    \label{eq:e0definition}
    \boldsymbol{n} \cdot \boldsymbol{q'} = - e  (\boldsymbol{n} \cdot \boldsymbol{q})
\end{equation} 
\begin{equation}
    \label{eq:m0definition}
    |\boldsymbol{n} \times \boldsymbol{J}| = - \mu  (\boldsymbol{n} \cdot \boldsymbol{J})
\end{equation}
\begin{equation}
 \boldsymbol{n} \times \boldsymbol{q'} = - \beta  (\boldsymbol{n} \times \boldsymbol{q}) \label{eq:beto0defintion}    
\end{equation}
These collision parameters define the instantaneous collision and are an input of the hard-sphere collision model.
The values are taken from \citet{Bunke2024a} for dry particles in the validation and set to 1 for the parametric study. 
Additionally, these coefficients are evaluated over the particle-particle interaction, using the extrapolation of position and velocity at the moment of bridge formation to the collision and the extrapolation of these variables from the bridge rupture to the collision.
The normal restitution coefficient is of primary interest for describing the head on collisions studied in this work. 
For further information, the reader is referred to our previous works \cite{Bunke2024a, huijgendrywetcollision}. \\ \\
Similar to the work of \citet{huijgendrywetcollision}, the simulations consist of two stages. 
In the first stage, the particles are placed in the computational domain and move in the domain with their prescribed initial velocity while the liquid layer surrounding the particle develops. 
To prevent the particles from leaving the computational domain or colliding, a window shifting algorithm is employed.
In this algorithm, the domain is divided into two regions, each containing a single particle. 
When the particle moves beyond the prescribed distance, of one grid cell, from its original position, both the particle and the fields in its region are moved back to this original position. 
The boundary of the two regions is located at L/2 and H/2, coinciding with the expected collision plane.
Once the liquid layers have fully developed and reached a stable configuration, the particles are released. 
Their motion is subsequently governed by Eq. \ref{eq::TranslationBalanseModified}, allowing them to approach each other and collide. 
During this stage, the window-shifting algorithm is disabled.
Collision data are post-processed using a custom Python script, while visualizations are generated using the PyVista toolbox \cite{Pyvista}.


\section{Experimental methods}
\begin{figure}
    \centering
    \includegraphics[width=0.85\linewidth]{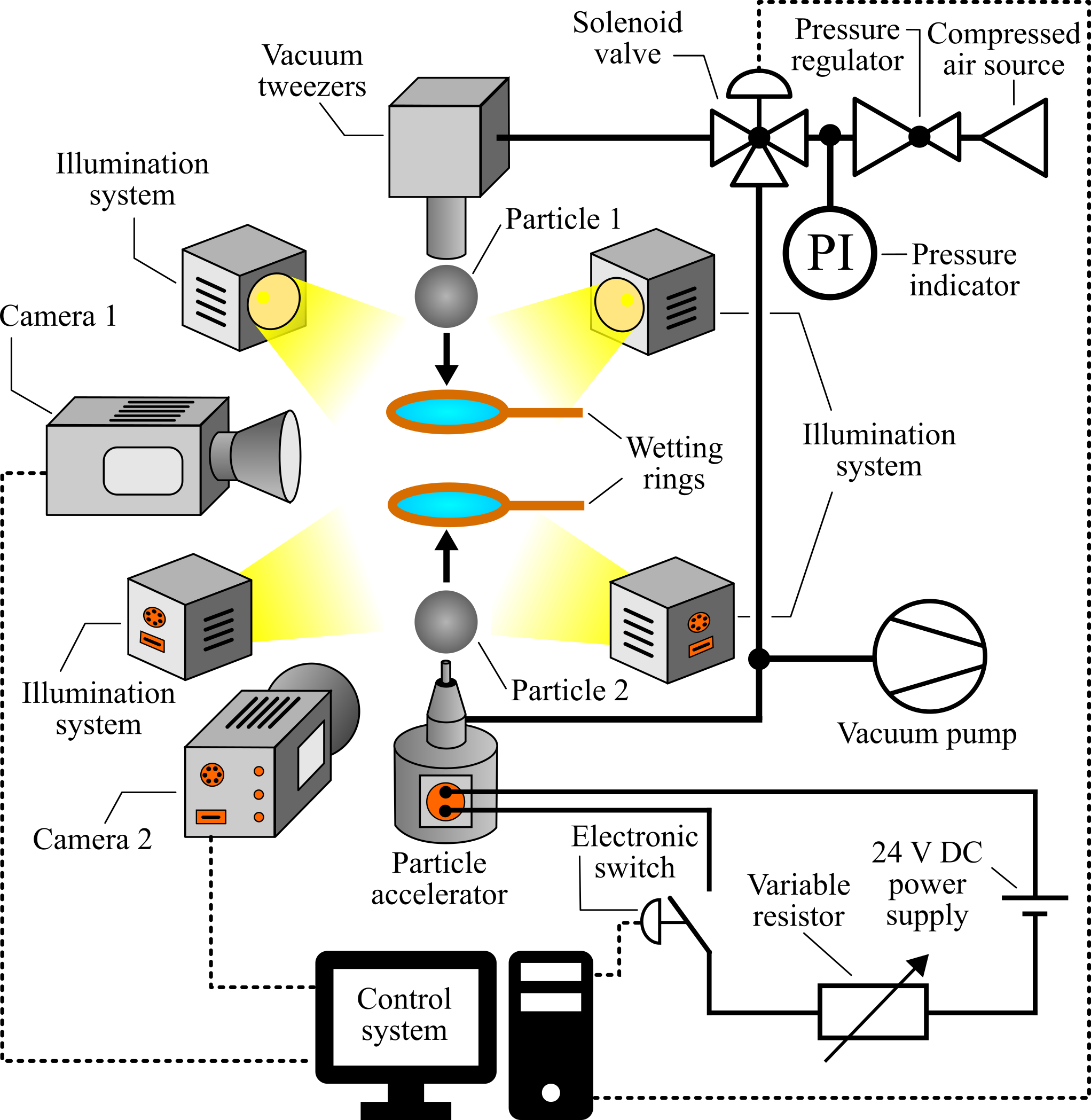}
    \caption{Experimental setup.}
    \label{fig::Experimental::setupoverview}
\end{figure}

\begin{figure}
    \centering
   \includegraphics[width=0.45\linewidth]{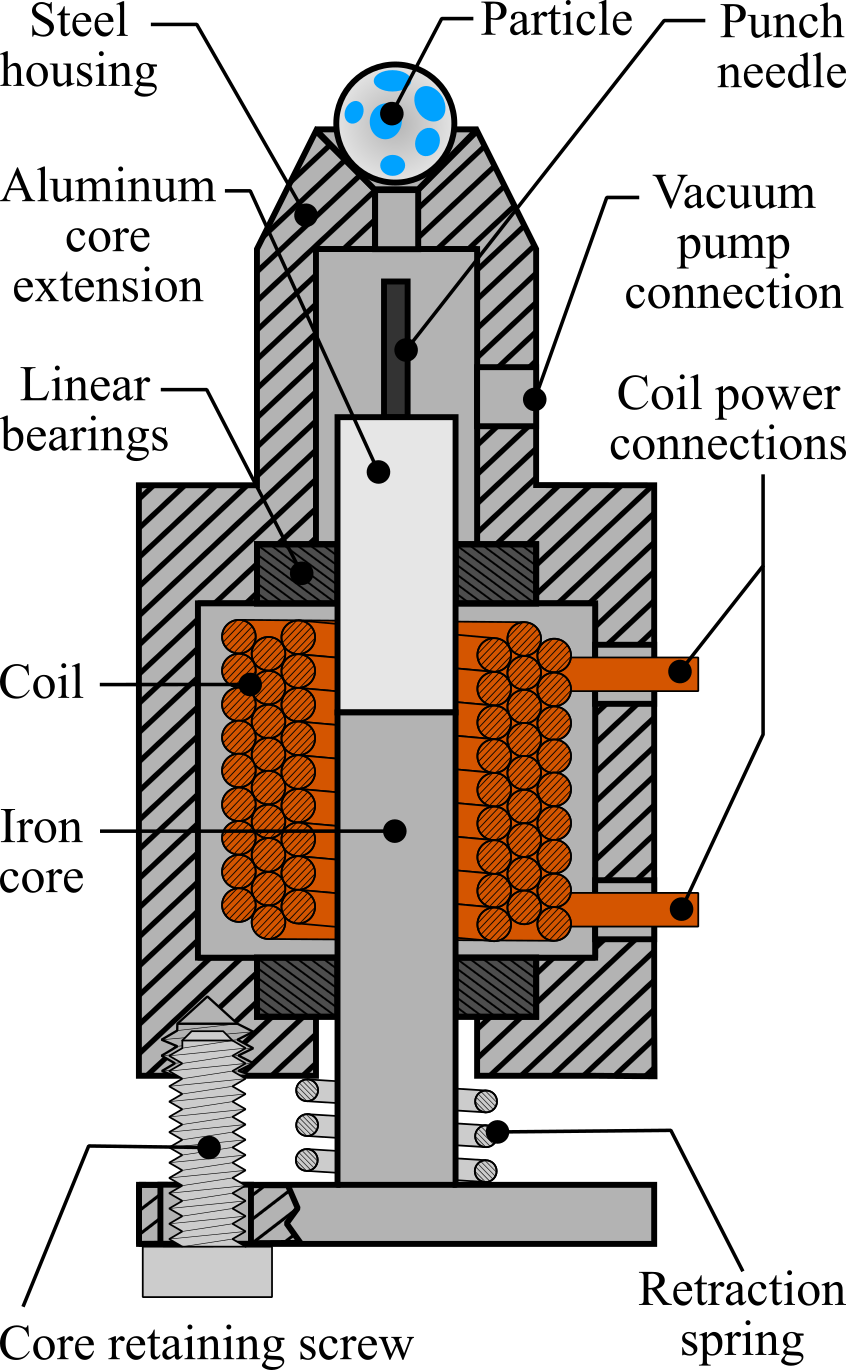}
   \caption{Schematic cross-section of the particle accelerator.}\label{fig::Experimental::plungerdesign}
\end{figure}

The experimental setup is adapted from the method described by \citet{Bunke2024a} for conducting and analyzing free collisions between two spherical particles. The basic principle of the setup is illustrated in Figure \ref{fig::Experimental::setupoverview}. The first particle is held in the upper position by a vacuum tweezers. The control system switches solenoid valves, which interrupt the connection between the vacuum tweezers and the vacuum pump while simultaneously applying a weak, brief overpressure pulse to the particle, causing it to detach from the tweezers at a defined time and fall. The second particle is launched from the bottom upward to hit the falling particle. This is achieved using the so-called particle accelerator. The schematic design of the particle accelerator is illustrated in Figure \ref{fig::Experimental::plungerdesign}. The particle is held in place on a hollow tip by an applied vacuum. A flat-tipped punch needle is positioned beneath the particle. This needle is attached to an iron core, which is placed eccentrically within a copper wire coil. The control system closes the electrical circuit, whereupon the iron core is drawn into the current-carrying coil by the resulting magnetic field. The needle attached to the iron core propels the particle upward, countering the falling particle. The punch and, consequently, the particle velocity can be adjusted via a variable resistor. \\
\begin{figure}
   \centering
   \includegraphics[width=1\linewidth]{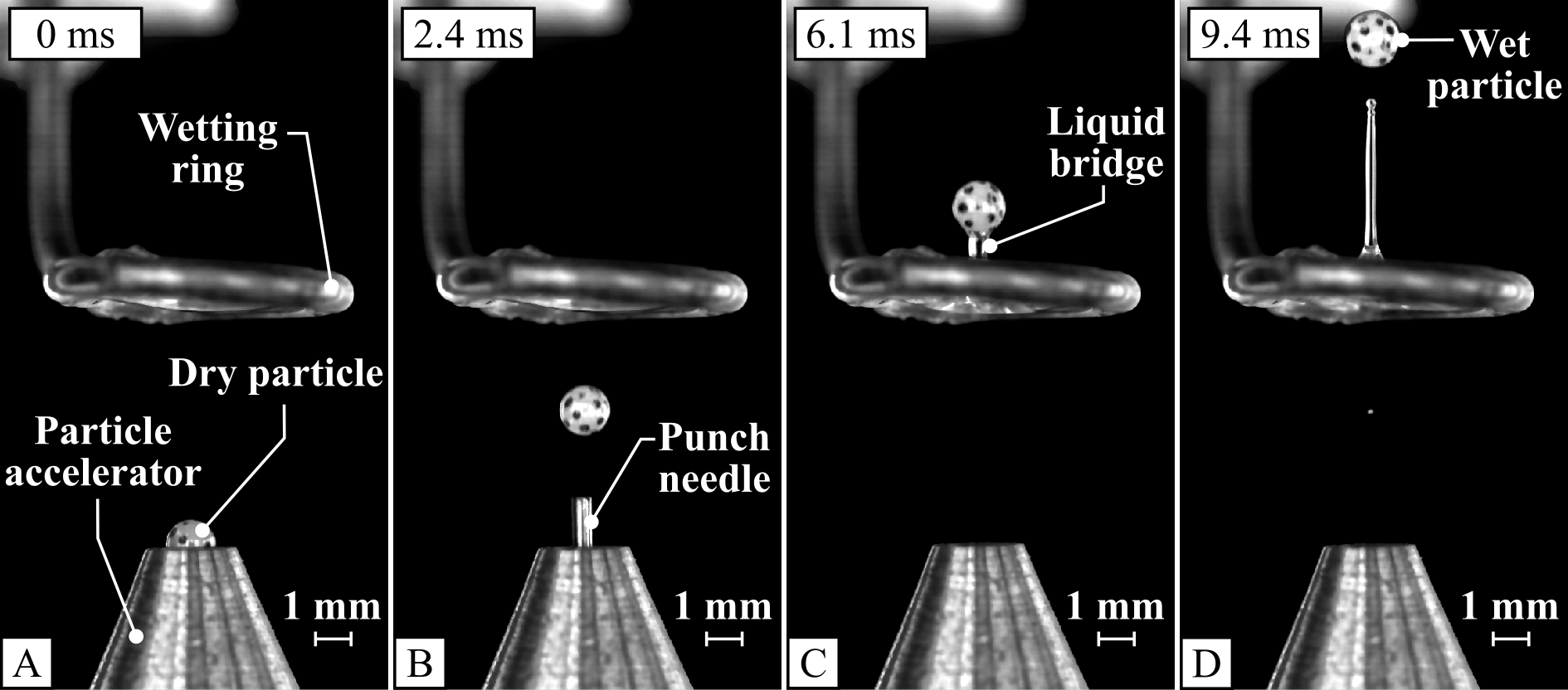}
   \caption{Operating principle of the particle accelerator and the particle wetting process.}\label{fig::Experimental::plungerwet}
\end{figure}

The punch needle approach used here for the particle accelerator differs fundamentally from the compressed-air-driven particle accelerator used by \citet{Bunke2024a, BUNKE2026122951}. The reason for this is the particle wetting process. Behind the vacuum tweezers and the particle accelerator, in the path of the particles, there is a wetting ring in each case. These rings serve to completely wet the particles with a liquid before the particles collide with one another. The wetting rings consist of round copper wire. The wire diameter is 1 mm, and the inner diameter of the ring is 5 mm. Before the experiment begins, these rings are briefly immersed in the liquid intended to wet the particles. This causes a liquid film to form inside the ring. When the particles pass through this film, they become completely wetted. This wetting process, as well as the operation of the particle accelerator, is illustrated in Figure \ref{fig::Experimental::plungerwet}. Although it is possible to accelerate the lower particle with compressed air, as in \citet{Bunke2024a, BUNKE2026122951}, the accelerating air flowing partially ahead of the particle can deform the liquid film in the wetting ring, which can lead to deflection of the particle during the wetting process due to asymmetric forces. In extreme cases, the liquid layer in the ring may even be destroyed before the particle can be wetted. The punch needle design used here prevents deformation and cracks in the liquid layer of the wetting ring, which significantly improves firing accuracy and the reliability of the wetting process. 

The particle accelerator and vacuum tweezers are aligned with each other using positioning systems, and their activation is coordinated by the control system so that the wetted particles collide within the field of view of two synchronized high-speed cameras. The positioning system allows the angle of the collision to be adjusted from a direct head-on collision to an extremely oblique one. The cameras used are \textit{CrashCam MINI HD} high-speed cameras manufactured by Integrated Design Tools Inc. (IDT), USA. The cameras are capable of 36-bit color recording and have a maximum image resolution of 1920 x 1080 pixels at 5,850 frames per second at full resolution. The experiments are recorded with a slightly reduced image width, which enables and utilizes a higher frame rate of 6,000 frames per second. To ensure sufficient illumination, the cameras are synchronized with up to four pulsed high-power LED lights from the \textit{miniConstellation 120} series (IDT, USA), each of which delivers up to 24,000 lumens. The camera and illumination system are thus significantly more powerful than the systems used in the studies by \citet{Bunke2024a, BUNKE2026122951}, which positively influences the accuracy of the measurements. \\

The cameras are arranged with a 90\textdegree~horizontal offset to capture the motion of the particles in all spatial directions. Based on the camera recordings, the translational motion and rotation of the particles in space, as well as the amount of liquid on the particles, are determined. To determine the rotation, the particles are marked with point-shaped rotation markers. The image analysis methods used are described in detail in the study by \citet{Bunke2024a}. \\

The particles used are the same as those used in our previous work \cite{huijgendrywetcollision, Bunke2024a, BUNKE2026122951}. These are highly spherical ball bearings made of yttrium-stabilized zirconia ($ZrO_2$) with a diameter of 1.5 $mm$. The spheres have a maximum roundness deviation of 0.13 $\mu m$, a density of 6050 $kg/m^3$, a Young modulus of 210 $GPa$, and a hardness of 89 $\pm$ 2 $HRA$. The surface of the particles is factory-polished and very smooth, with surface roughness values of 0.026 $\pm$ 0.003 $\mu m$ (arithmetic mean roughness -- Ra) and 0.198 $\pm$ 0.045 $\mu m$ (mean roughness depth -- Rz) \cite{BUNKE2026122951}. With a dry normal restitution coefficient, which, at 0.998 (particle-particle) and 0.984 $\pm$ 0.022 (particle-plate) \citet{Bunke2024a}, is very close to 1, these are excellent model particles for nearly ideal elastic collision behavior. \\

The liquid used to wet the particles is silicone oil. The reason for using silicone oil is that, unlike water, for example, it does not evaporate or dry out significantly during the experiment. A total of three different silicone oils with varying viscosities are used. The physical properties of the silicone oils used are listed in Table \ref{tab::experiment::viscousoils}. Both particles were always wetted with the same silicone oil in each experimental run. 
\begin{table}[]
\caption{Properties of silicone oil at 20 \textdegree C.}\label{tab::experiment::viscousoils}
\scalebox{0.9}{%
\begin{tabular}{cccc}
\hline
Silicone oil Nr. & Viscosity ($mPa\cdot s$) & Surface tension ($mN/m$) & Density ($kg/m^3$) \\
\hline
1 & $21.11 \pm 0.09$ & $20.17 \pm 0.02$ & $954.06 \pm 1.02 $  \\
2 & $56.57 \pm 0.15$ & $20.30 \pm 0.01$ & $961.74 \pm 0.87$   \\
3 & $107.34 \pm 0.61$ & $21.03 \pm 0.02$ & $970.04 \pm 1.81 $ \\
\hline
\end{tabular}}
\end{table}
\section{Verification and Validation}
The numerical framework used in this study was previously verified in \citet{huijgendrywetcollision}. 
The framework was validated for dry-wet collisions using an viscous oil with a viscosity of 18.9 $mPa\cdot s$. 
In this work, wet-wet collisions using Silicone oil Nr. 3 are additionally considered.
Validation with different fluid properties and initial conditions increases confidence in the framework. The initial conditions are taken directly from the experiment and listed in Table \ref{tab:validaiton:settingsWetcollision}.\\
\begin{table}[]
\caption{Simulation settings validation}\label{tab:validaiton:settingsWetcollision}
\scalebox{0.9}{%
\small\addtolength{\tabcolsep}{-2pt}
\begin{tabular}{|l|cll|l|}
\hline
Parameter      & \multicolumn{1}{l|}{Case 1} & \multicolumn{1}{l|}{Case 2}  & Case 3 & Unit     \\ \hline
Collision angle     & \multicolumn{1}{l|}{175.0}  & \multicolumn{1}{l|}{146.5}   & 136.1  & \textdegree   \\ \hline
liquid layer ($h_1, h_2$)   & \multicolumn{1}{l|}{127.3, 75.2}   & \multicolumn{1}{l|}{55.4, 99}       & 86, 102.8  & $\mu m$ \\ \hline
velocity P1 ($v_1$)  & \multicolumn{1}{l|}{-0.011, 0.015, -0.94} & \multicolumn{1}{l|}{-0.006, 0.015, -1.04} & 0.017	-0.0064	-0.98  & $m/s$      \\ \hline
velocity P2 ($v_2$)    & \multicolumn{1}{l|}{0.019, 0.022, 0.85}   & \multicolumn{1}{l|}{0.030 0.021 0.85}      & 0.016, -0.006, -0.98   & $m/s$      \\ \hline
angular velocity P1 ($\omega_1$)& \multicolumn{1}{l|}{1.52, -77.1, -3.4}    & \multicolumn{1}{l|}{45.9, 57.8, 1.4}       &    0.035, 0.031, 0.83 & $1/s$      \\ \hline
angular velocity P2 ($\omega_2$)& \multicolumn{1}{l|}{-29.2, 145.3, -3.0}   & \multicolumn{1}{l|}{-73.2, 129.1, 5.87}    &  -59.6, 171.2, 0.0  & $1/s$      \\ \hline
Particle diameter ($D_p$)       & \multicolumn{3}{c|}{1.5}  & $mm$       \\ \hline
Resolution     & \multicolumn{3}{c|}{30}   & $D_p/\Delta x$    \\ \hline
L,H,D   & \multicolumn{3}{c|}{6$D_p$, 6$D_p$, 6$D_p$}      & $[-]$  \\ \hline
air viscosity ($\mu_g$)  & \multicolumn{3}{c|}{0.018365}   & $mPa\cdot s$     \\ \hline
liquid viscosity ($\mu_l$)   & \multicolumn{3}{c|}{100}  & $mPa\cdot s$     \\ \hline
air density ($\rho_g$)    & \multicolumn{3}{c|}{1.1845}      & $kg/m^3$    \\ \hline
liquid density ($\rho_l$)       & \multicolumn{3}{c|}{{968}}       & $kg/m^3$    \\ \hline
solid density ($\rho_s$) & \multicolumn{3}{c|}{6050} & $kg/m^3$ \\ \hline
surface tension ($\sigma$)     & \multicolumn{3}{c|}{21}   & $mN/m$     \\ \hline
$e_0$, $\beta_0$, $\mu_0$     & \multicolumn{3}{c|}{0.998, 0.406, 0.077}      & $[-]$  \\ \hline
\end{tabular}}
\end{table}

\begin{figure}[]
    \centering
    \begin{minipage}[c]{0.48\textwidth}
 \centering
 \begin{subfigure}{\textwidth}
   \includegraphics[width=\textwidth]{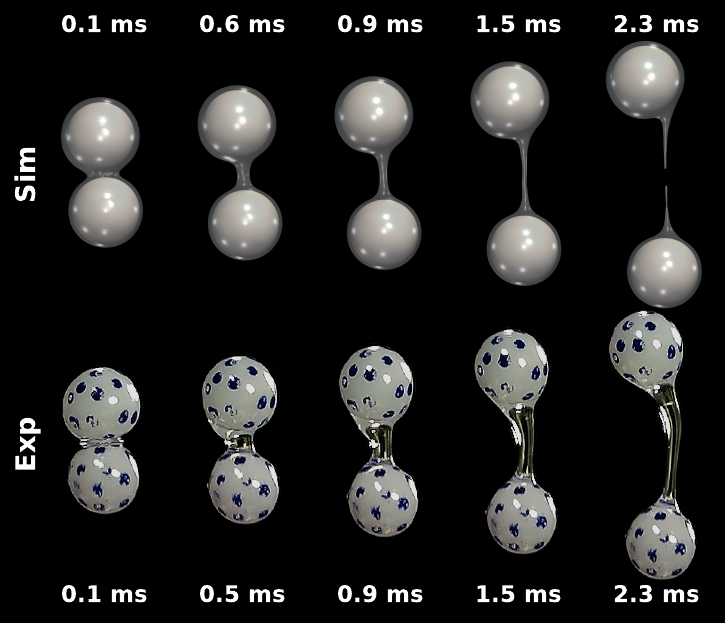}
   \caption{Case 1: Collision angle = 175.0\textdegree}\label{fig::validation::case01::sidebyside}
 \end{subfigure}
    \end{minipage}
    \hfill
    \begin{minipage}[c]{0.48\textwidth}
 \centering
 \begin{subfigure}{\textwidth}
   \includegraphics[width=\textwidth]{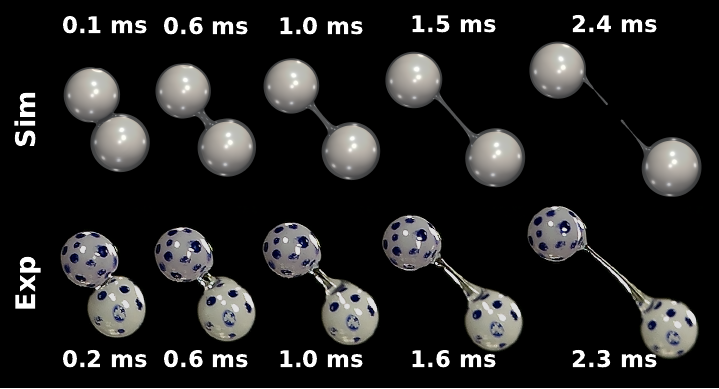}
   \caption{Case 2: Collision angle = 146.5\textdegree}\label{fig::validation::case30::sidebyside}
 \end{subfigure}
 \vspace{0.5em}
 \begin{subfigure}{\textwidth}
   \includegraphics[width=\textwidth]{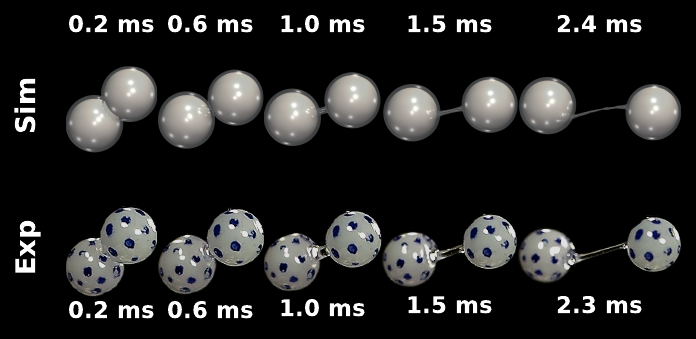}
   \caption{Case 3: Collision angle = 136.1\textdegree}\label{fig::validation::case44::sidebyside}
 \end{subfigure}
    \end{minipage}
    \caption{Qualitative comparison of the formed particle liquid bridge complex after collision for both experiments (Exp) and simulations (Sim) for three different collision angles (175.0\textdegree, 146.5\textdegree, 136.1\textdegree) }\label{fig::validation::sidebyside}
\end{figure}

\begin{figure}
    \centering
 \begin{subfigure}[b]{0.48\textwidth}
   \centering
   \includegraphics[width=\textwidth]{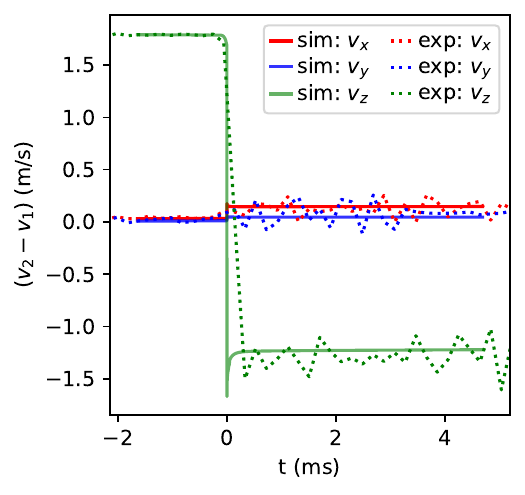}
   \caption{Case 1: Collision angle = 175.0\textdegree}\label{fig::validation::case01::velocities}
    \end{subfigure}
     \begin{subfigure}[b]{0.48\textwidth}
   \centering
   \includegraphics[width=\textwidth]{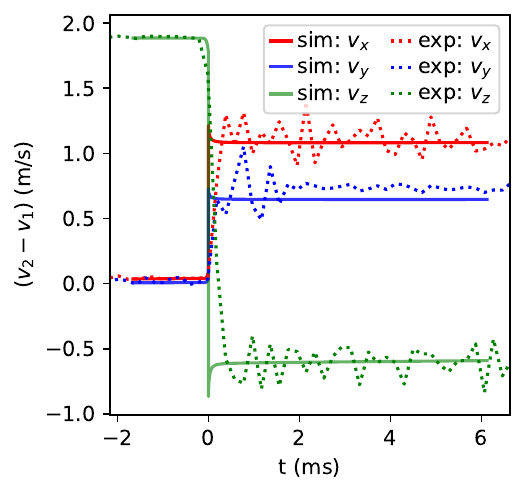}
   \caption{Case 2: Collision angle = 146.5\textdegree}\label{fig::validation::case30::velocities}
    \end{subfigure}
     \begin{subfigure}[b]{0.48\textwidth}
   \centering
   \includegraphics[width=\textwidth]{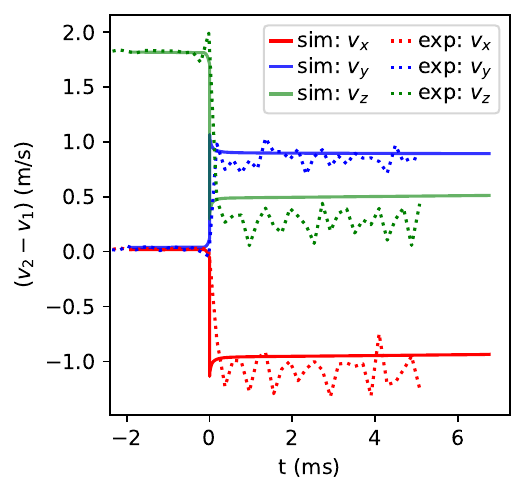}
   \caption{Case 3: Collision angle = 136.1\textdegree}\label{fig::validation::case44::velocities}
    \end{subfigure}
  
    \caption{Quantitative comparison of the translational velocity for both experiments and simulations for three different collision angles (175.0\textdegree, 146.5\textdegree, 136.1\textdegree)}\label{fig::validation::velocities}
\end{figure}

The first case considers a nearly head-on collision (Collision angle = 175.0\textdegree) where the particles have a substantial initial rotation orthogonal to the impact velocity. 
Figure \ref{fig::validation::case01::sidebyside} 
and Video \ref{vid:case01} demonstrate good agreement between the simulation and the experiment. 
Both experiment and simulation show that rotation displaces the liquid bridge, consistent with \citet{BUCK2017218}. 
The liquid bridge ruptures earlier in the simulations and this is mainly dependent on the grid resolution used.
However, the final stages of liquid bridge extension do not significantly affect the rebound velocity \cite{huijgendrywetcollision}. 
At t = 2.3 ms, the rotation of the bridge attachment agrees closely between the experiment and the simulation, indicating accurate prediction of the angular velocity. 
Figure \ref{fig::validation::case01::velocities} shows good agreement in translational velocities. The figure clearly shows that the velocity measurements exhibit oscillations due to particle detection in the presence of the liquid bridge.
However, the simulation results remain within this measurement range.
These observations demonstrate that the framework captures the governing physics and can be employed for further explorations. 
This case is particularly suitable for analyzing the normal restitution coefficient as the collision is head-on.\\

Validation for oblique collisions, while not essential for deriving normal restitution coefficients, demonstrates the capabilities of the method.
Figure \ref{fig::validation::case30::sidebyside} and Video \ref{vid:case30} demonstrate good agreement between experiments and simulations for an collision angle of 146.5\textdegree. 
As in the head-on case, liquid bridge rupture occurs earlier in the simulations. 
While the rupture of the liquid bridge is numerically governed by the grid resolution, the experimental visualization of these thin structures also introduces uncertainties.
Figure \ref{fig::validation::case30::velocities} shows that simulated translational velocities fall within the experimental range, with oscillations similar to those observed in Figure \ref{fig::validation::case01::velocities}. \\

The final case corresponds to the most oblique collision angle (136.1\textdegree). 
In this configuration, the normal of the collision plane is aligned with the xy-direction, such that the impact velocity is primarily tangential (z-direction) and the tangential motion is largely converted into normal slip velocity (xy-direction). 
This is observed in Figure \ref{fig::validation::case44::sidebyside} and Video \ref{vid:case44}, where the particle complex moves in the horizontal (x) direction and to the camera. Figure \ref{fig::validation::case44::velocities} confirms that the dominant components of the velocity after the collision are the x- and y-component. 
Experimental and simulated velocities show overlap. However, deviations increase for smaller velocity components (z rebound velocity). This is consistent with the drift in restitution coefficient observed by \citet{huijgendrywetcollision} and \citet{Bunke2024a} at lower collision angles. 
This indicates that highly oblique collisions are less suitable for determining normal restitution coefficients. 
Therefore, only case 1 is considered in the subsequent analysis of the normal restitution coefficient.

\section{Results}
\subsection{Parametric study}
The validated framework is used to investigate a range of conditions based on validation case 1 from Table \ref{tab:validaiton:settingsWetcollision}.
In the base case simulation, the liquid layer thickness of 0.05 $D_p$ on each particle is used corresponding to a liquid content of 66.2\%. 
The collision angle is 180\textdegree, and initially rotation free, gravity is neglected, and the dry restitution coefficient ($e_0$) is 1.0. 
The obtained wet normal restitution coefficient is thus normalized by the dry restitution coefficient. 
The investigated parameters are liquid viscosity ($25 - 600~mPa\cdot s$), liquid density ($200 - 1400~kg/m^3$), slip velocity ($0.4 - 4.0~m/s$), solid density ($1500 - 6050~kg/m^3$), liquid film thickness ($25 - 129~ \%~ V_{liquid}/V_{particle}$), and surface tension coefficient ($5-40~mN/m$). 

Note that under these conditions, the variations in surface tension do not significantly affect the restitution coefficient, which is also indicated by the dominance of the viscous forces over the surface tension forces (base case Capillary number is equal to 9.5).
This has also been reported in literature \cite{SHAO2022117090, PAN2026123656, Buck201821}, where it is shown that the surface tension primarily influences liquid-bridge morphology and rupture dynamics rather than the restitution coefficient. 
In addition, the energy dissipation via surface energy is negligible.
The effects of surface tension become relevant at low impact velocities, where inertia and viscous forces are relatively smaller \cite{SHAO2022117090, PAN2026123656, Buck201821}. 
The surface tension coefficient is thus not considered a dominant parameter for the collision dynamics investigated here.

\begin{figure}
    \centering
    \begin{subfigure}[b]{0.45\textwidth}
   \centering
   \includegraphics[width=\textwidth]{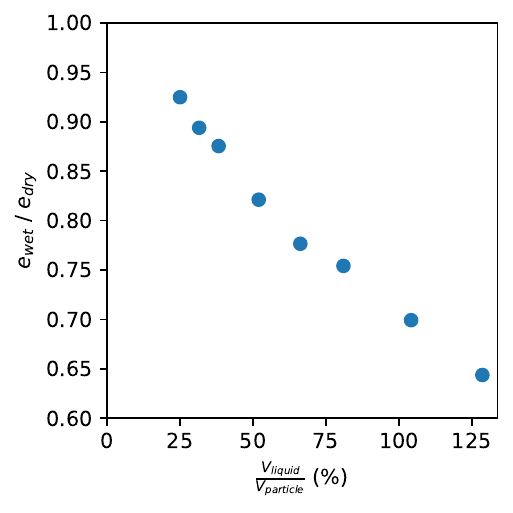}
   \caption{Layer thickness}\label{fig::results::paramstudy::thickness}
    \end{subfigure}
 \begin{subfigure}[b]{0.45\textwidth}
   \centering
   \includegraphics[width=\textwidth]{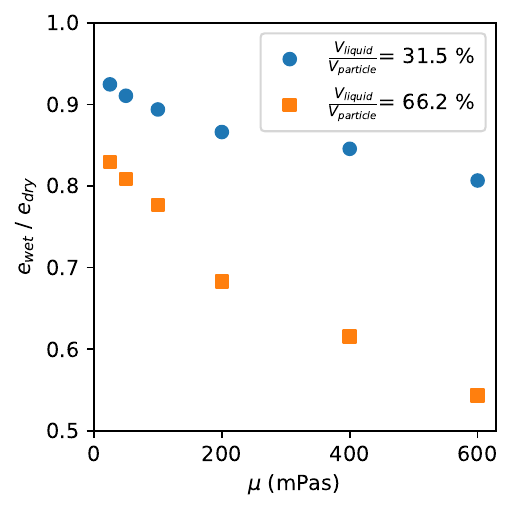}
   \caption{Viscosity}\label{fig::results::paramstudy::viscosity}
    \end{subfigure}
     \begin{subfigure}[b]{0.45\textwidth}
   \centering
   \includegraphics[width=\textwidth]{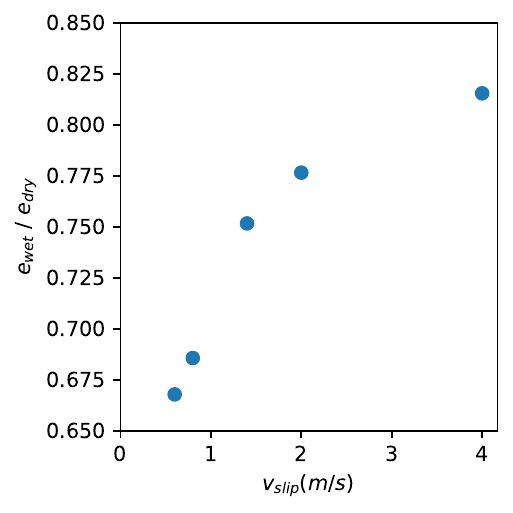}
   \caption{Slip velocity}\label{fig::results::paramstudy::velocity}
    \end{subfigure}     
    \begin{subfigure}[b]{0.45\textwidth}
   \centering
   \includegraphics[width=\textwidth]{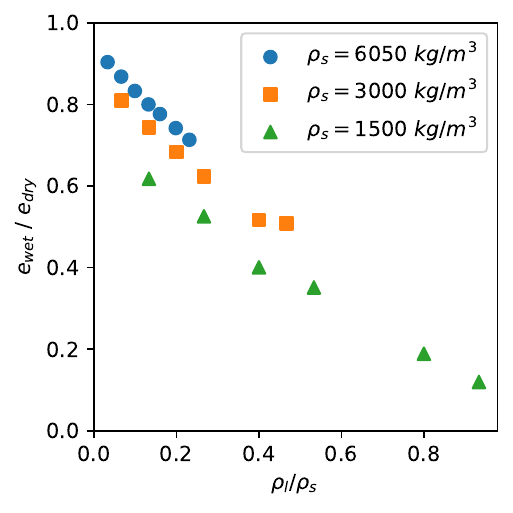}
   \caption{Density ratio}\label{fig::results::paramstudy::density}
    \end{subfigure}

    \caption{Effect of varying viscosity, density, velocity and liquid film thickness on the restitution coefficient for wet-wet collisions using simulations. The used base case has a liquid layer thickness of 0.1$D_p$, corresponding to a liquid content of 66\%, a liquid viscosity of 100 $mPa\cdot s$, a slip velocity of 2 $m/s$, a surface tension coefficient of 21 mN/m, a solid density of 6050 $kg/m^3$, and a liquid density of 968 $kg/m^3$.}\label{fig::results::paramstudy}
\end{figure}

Figure \ref{fig::results::paramstudy::thickness} shows a decrease of the  restitution coefficient with increasing liquid content. 
Similar trends have been reported in the literature for particle-wall, partially-wet particle, wet-wet particle, and wet-dry particle collisions \cite{WANG2025120546, FAN2024119308, SHAO2022117090, PAN2026123656, Buck201821, Buck2019CollisionExperiments, Bunke2024a, Antonyuk2009InfluenceImpact}.
There are two major effects of the liquid content on the collision dynamics: 
First, the particles must penetrate a thicker liquid layer before the particles make contact.
Second, a larger liquid bridge forms during the rebound, resulting in increased viscous dissipation.
Previous works have identified the latter mechanism as the dominant source of energy dissipation \cite{huijgendrywetcollision, WANG2025120546}. 
Its contribution also increases with the liquid content as the post-collision dissipation grows more rapidly than the pre-collision dissipation in the simulations.

A decrease in the restitution coefficient with increasing viscosity has been reported for a wide range of geometries \cite{Buck201821, Buck2019CollisionExperiments, WANG2025120546, Antonyuk2009InfluenceImpact,  FAN2025121253, PAN2026123656, SHAO2022117090}. 
Figure \ref{fig::results::paramstudy::viscosity} confirms this trend. 
Prior to the collision, liquid is squeezed out of the gap between the particles, whereas during the rebound liquid is drawn back into the expanding gap, as was already illustrated in \citet{huijgendrywetcollision}. 
In both stages, the dominant effect is viscous dissipation and therefore depends on the viscosity of the liquid \cite{PAN2026123656, Kan2015}. 
In the limit of zero viscosity ($\mu_l\rightarrow0$), the restitution coefficient should approach unity. 
This is particularly evident for the low liquid content, where the restitution coefficient rapidly approaches unity as the viscosity decreases.

Figure \ref{fig::results::paramstudy::velocity} shows that the restitution coefficient decreases with decreasing slip velocity, which has also been observed in literature for particle-wall, partially-wet particle, and wet-wet particle collisions \cite{Buck201821, Buck2019CollisionExperiments, FAN2024119308,  PAN2026123656, FAN2025121253, WANG2025120546, Kan2015, DANCZYK2022117112}. 
Increasing impact velocity increases particle inertia, thereby reducing the relative influence of viscous dissipation.
When viscous dissipation exceeds the initial kinetic energy, rebound is prevented and agglomeration occurs \cite{FAN2025121253}. 

Finally, the effect of density is quantified. 
In contrast to previously discussed parameters, density effects have received relatively little attention due to the limited availability of particles and liquids with different densities. 
Numerical simulations therefore provide a valuable tool for investigating these effects of density.
Figure \ref{fig::results::paramstudy::density} indicates a decrease in the restitution coefficient with increasing liquid density and decreasing solid density. 
Decreasing solid density reduces the particle inertia, similar to the effect of initial velocity, and results in a decrease of the restitution coefficient. 
Increasing the density of the liquid increases the inertia of the liquid, thus increasing the resistance of the liquid during draining and refilling the gap.
These two effects can be combined when considering the liquid-to-solid density ratio, showing and decrease of the restitution coefficient with a increase in the liquid-solid density ratio. 
\subsection{Stokes model}\label{sec::results::stokesmodel}
\begin{figure}
    \centering
    \begin{subfigure}[b]{0.48\textwidth}
   \centering
   \includegraphics[height=0.82\textwidth,left]{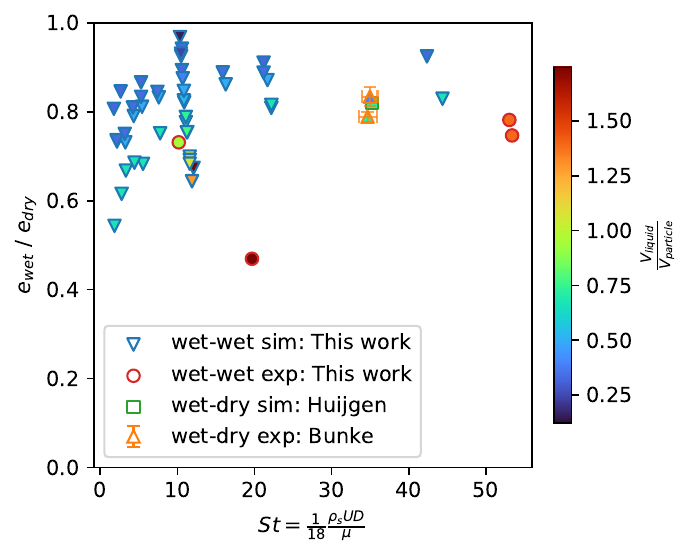}
   \caption{$St_{Davis}$ (Eq. \ref{eq::StokesNumberDavis})}\label{fig::results::StokesNumbers1::StokesDavis}
    \end{subfigure}
  \begin{subfigure}[b]{0.48\textwidth}
   \centering
   \includegraphics[height=0.82\textwidth,left]{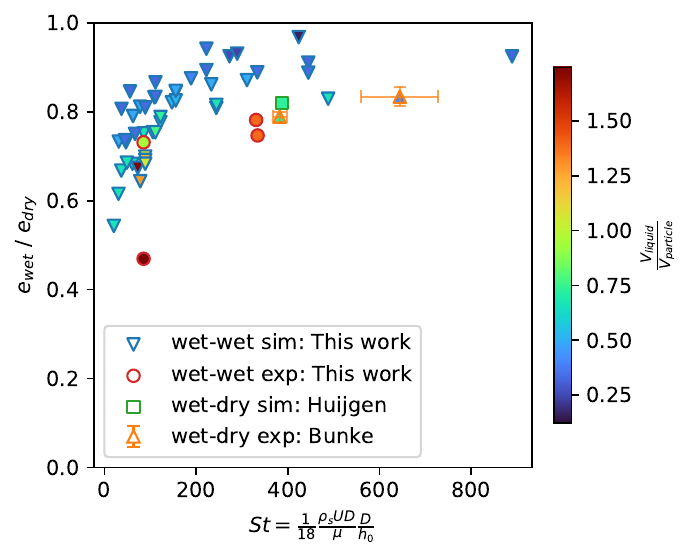}
   \caption{$St_{Ma}$ (Eq. \ref{eq::StokesNumberMa})}\label{fig::results::StokesNumbers1::StokesMa}
    \end{subfigure}  
    \begin{subfigure}[b]{0.48\textwidth}
   \centering
   \includegraphics[height=0.82\textwidth,left]{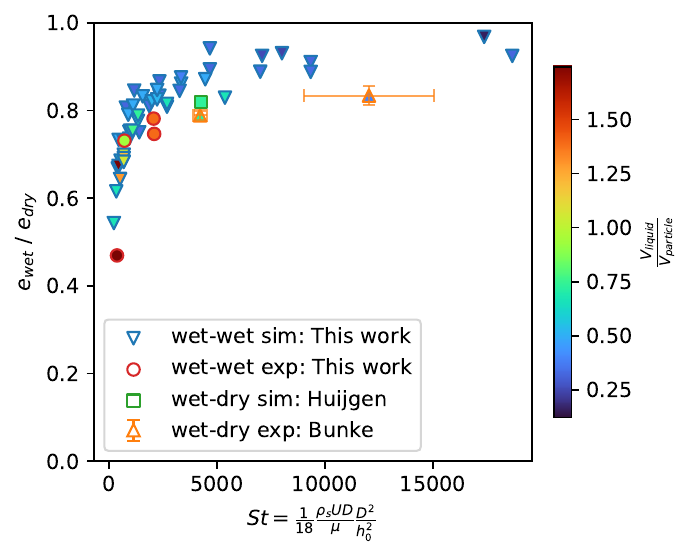}
   \caption{$St_{visc}$ (Eq. \ref{eq::StokesNumberNewVisc})}\label{fig::results::StokesNumbers1::StokesNewVisc}
    \end{subfigure}  
 \begin{subfigure}[b]{0.48\textwidth}
   \centering
   \includegraphics[height=0.82\textwidth,left]{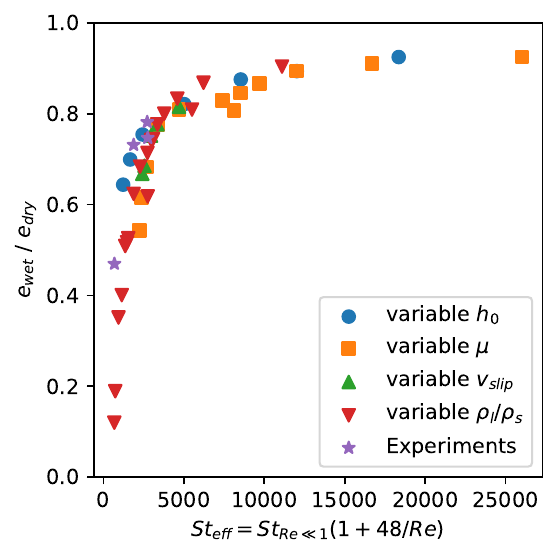}
   \caption{$St_{eff}$ (Eq. \ref{eq::resutls::newStokesNumberViscInertia})}\label{fig::results::StokesNumbers2::StokesNew}
    \end{subfigure}  
    \caption{Restitution coefficient for wet-wet collisions at various viscosities, velocities and liquid content vs Stokes number definitions. Figure a,b,c show the current simulations and experiments for wet-wet collisions, and previous works on wet-dry collisions \cite{huijgendrywetcollision, Bunke2024a}. Figure d shows the comparison of the current experiments with the simulations, indicating the different trends from the parameter study.}\label{fig::results::StokesNumbers1}
\end{figure}

The objective of this study is to unify these effects within a single predictive model. A common approach is to combine the physical properties into a non-dimensional number. 
\citet{FAN2025121253} used Weber and Reynolds numbers to quantify the agglomeration behavior.
For wet particle collisions, the Stokes number is one of the most commonly used dimensionless groups \cite{DAVIS_RAGER_GOOD_2002}. 
The Stokes number is defined as the ratio of the characteristic particle response time to the characteristic fluid time scale. 
In wet particle collisions, the dominant mechanism is the draining and refilling of the gap between the particles with liquid.
For these small particles, gravity is negligible and drag dominates, so the force balance on the particle reduces to:
\begin{equation}\label{eq::StokesDerivation::forcebalance}
    m_p\frac{dU}{dt}=-F_D= -\frac{1}{2}\rho_l U^2C_dA_p 
\end{equation}
Under creeping-flow conditions ($Re \ll 1$), the drag coefficient reduces to $C_d=24/Re$. The Reynolds number is $Re = D\rho_lU_0/\mu_l$, with D as the complex diameter and $U_0$ the initial particle velocity. The resulting particle velocity is given by: 
\begin{equation}\label{eq::StokesDerivation::FinalODEViscousRange}
    U = U_0\exp\left(-\frac{18\mu}{D^2\rho_s}t\right) = U_0\exp\left(-\frac{t}{t_{solid, Re\ll1}}\right) 
\end{equation}
Which introduces the characteristic particle time scale ($t_{solid, Re\ll1}$):
\begin{equation}\label{eq::StokesDerivation::SolidTimeViscousRange}
    t_{solid, Re\ll1} = \frac{D^2\rho_s}{18\mu}
\end{equation}
The characteristic fluid time scale is commonly defined as the characteristic length scale divided by the characteristic fluid velocity.
For the characteristic fluid time scale ($t_{fluid}$) we use $D/U_0$, or written in dimensionless form:
\begin{equation}\label{eq::StokesNumberDavis}
    St_{Davis}=\frac{t_{solid, Re\ll1}}{t_{fluid}} = \frac{\rho_sU_0D}{18\mu}
\end{equation}
\citet{DAVIS_RAGER_GOOD_2002} proposed correlating the restitution coefficient with this Stokes number (Eq. \ref{eq::StokesNumberDavis}) in Equation \ref{eq::StokesModelDavis}. 
The critical Stokes number $St_{crit}$ is based on the elasticity parameter, which depends on properties of the liquid and particles and $e_\infty$ is the restitution coefficient for high Stokes numbers.

\begin{equation}\label{eq::StokesModelDavis}
    e_{wet}=e_{\infty}\left(1-\frac{St_{crit}}{St}  \right)
\end{equation}
Figure \ref{fig::results::StokesNumbers1::StokesDavis} shows that this Stokes number collapses the variations in viscosity and impact velocity. 
However, it does not capture the variations in liquid-layer thickness. 

As an alternative, \citet{Ma2013StokesNumber} hypothesized that the characteristic fluid length scale should be the liquid layer height $h_0$, describing the Stokes number in Eq. \ref{eq::StokesNumberMa}. 
Since then, this definition has been subsequently adopted in numerous studies \cite{Ma2016, FAN2025121253, BUCK2017218, FAN2024119308}. Note that this derivation is based on a particle-plate collision. It can be extended to wet particle-particle collisions, as the lubrication force is maximum in the near-center area of the spheres, while the pressure force diminishes for the convex area. To apply this to wet-wet collisions, the liquid layer on the two particles is summed.
\begin{equation}\label{eq::StokesNumberMa}
    St_{Ma}=\frac{t_{solid, Re\ll1}}{t_{fluid}} =\frac{\rho_sU_0D}{18\mu}\frac{D}{h_0}
\end{equation}
Figure \ref{fig::results::StokesNumbers1::StokesMa} shows that this definition incorporates the effects of the liquid-layer thickness partly, but there is still a clear dependence of the liquid-layer thickness on the results. 

\begin{figure}
    \centering
    \includegraphics[width=0.5\linewidth]{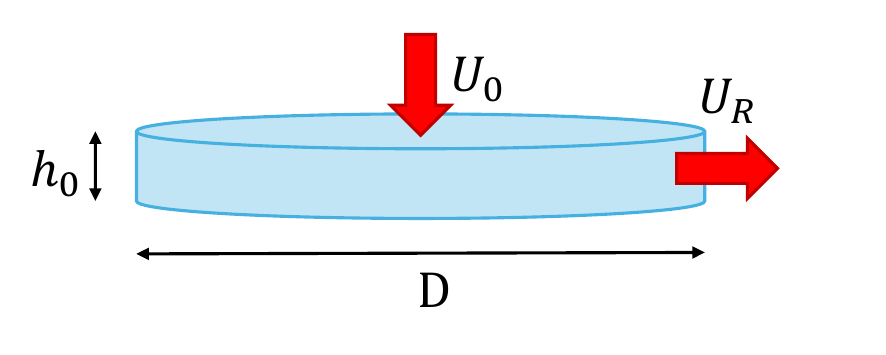}
    \caption{Sketch of liquid squeezed in radial direction. }
    \label{fig:squeezeoutliquidsketch}
\end{figure}

To ensure that all the effects of the liquid-layer thickness are included, we propose that the characteristic fluid velocity should be defined as the radial velocity associated with the liquid drainage from the gap between two particles, rather than the approach velocity $U_0$.
To determine this velocity, the liquid layer can be approximated as a cylindrical disk (Figure \ref{fig:squeezeoutliquidsketch}). This disk decreases in the longitudinal z-direction and increases in the radial direction when the particles approach.

Based on mass conservation, the flow induced at the bases of the cylinder should be equal to the flow created on the lateral surface:
\begin{equation}
   U_0 \pi D^2/4 = U_R \pi D h_0
\end{equation}
Note that the relative velocity of the particles is equal to $U_0$. This results in the following fluid velocity:
\begin{equation}
 U_R \propto \frac{D}{h_0}U_P
\end{equation}
 Based on this velocity, a new Stokes number $St_{visc}$ is derived as
 \begin{equation}\label{eq::StokesNumberNewVisc}
 St_{visc}=\frac{t_{solid, Re\ll1}}{t_{fluid}} = \frac{D^2\rho_s}{18\mu}\frac{U_{R}}{h_0} =\frac{D^2\rho_s}{18\mu}\frac{U_0D}{h_0^2} =\frac{\rho_sU_0D}{18\mu}\frac{D^2}{h_0}
\end{equation}
Figure \ref{fig::results::StokesNumbers1::StokesNewVisc} demonstrates that this definition successfully collapses the effects of viscosity, impact velocity, and liquid content. Additionally, it nicely overlaps with the performed experiments for all silicone oils. The exact conditions can be found in \ref{sec::appendix1::rawdata}. It should be noted that the experimental validation results show better agreement compared to the results of \citet{Bunke2024a}. This can be attributed the improvements made to the experimental set-up (i.e. improvements in the imaging, illumination, and process control), and the difference in the liquid coverage of the particles as in the works of \citet{Bunke2024a} and \citet{huijgendrywetcollision} only one particle is wet.

However, this definition of the Stokes number is not able to include the effects of density ratio, as it is not included in the original derivation. The reason for this is the approximation of $Re \ll 1$ in the derivation of the Stokes number, effectively neglecting any inertial effects. To account for the effects for inertia, we propose to use the limiting drag coefficient at high Reynolds numbers ($Re > 1000$, $C_D \approx 1/2$). Using this definition in Eq. \ref{eq::StokesDerivation::forcebalance}, results in:
\begin{equation}
    \frac{dU}{dt} = -\frac{3C_D}{4 D}\frac{\rho_l}{\rho_s}U^2 = -\frac{3}{8 D}\frac{\rho_l}{\rho_s}U^2
\end{equation}
When linearized, it shows the following solid time
\begin{align}
    \frac{dU}{dt} = -\frac{3}{8}\frac{U_0}{D}\frac{\rho_l}{\rho_s}U = \frac{1}{t_{solid, Re>1000}}U &  &
    t_{solid, Re>1000} = \frac{8}{3}\frac{\rho_s}{\rho_l} \frac{D}{U_0}
\end{align}
The inertial Stokes number can thus be defined as:
\begin{equation}\label{eq::StokesNumberNewInertia}
     St_{inertia}=\frac{t_{solid, Re>1000}}{t_{fluid}} = \frac{D U}{h_0^2} t_{solid, Re>1000} =  \frac{8}{3}\frac{\rho_s}{\rho_l}\frac{D^2}{h_0^2} = \frac{48}{Re} St_{visc}
\end{equation}
This $St_{inertia}$ does captures the influence of the density ratio, but the effects of the viscosity are not captured with this Stokes number. 
Ideally, these two Stokes numbers ($St_{visc}, St_{inertia}$) should be combined. 
Superimposing these two Stokes numbers give the following relation
\begin{equation}\label{eq::resutls::newStokesNumberViscInertia}
    St_{eff}=St_{visc} + St_{inertia} = St_{visc}\left(1+\frac{48}{Re}\right)  = \frac{U_0 D^3\rho_s}{18\mu h_0^2}\left(1+\frac{48}{Re}\right)
\end{equation}
Figure \ref{fig::results::StokesNumbers2::StokesNew} shows that this relation can effectively unify the effects of viscosity, density, slip velocity and liquid content. \\

\begin{figure}
    \centering 
    \begin{subfigure}[b]{0.48\textwidth}
   \centering
   \includegraphics[width=\textwidth]{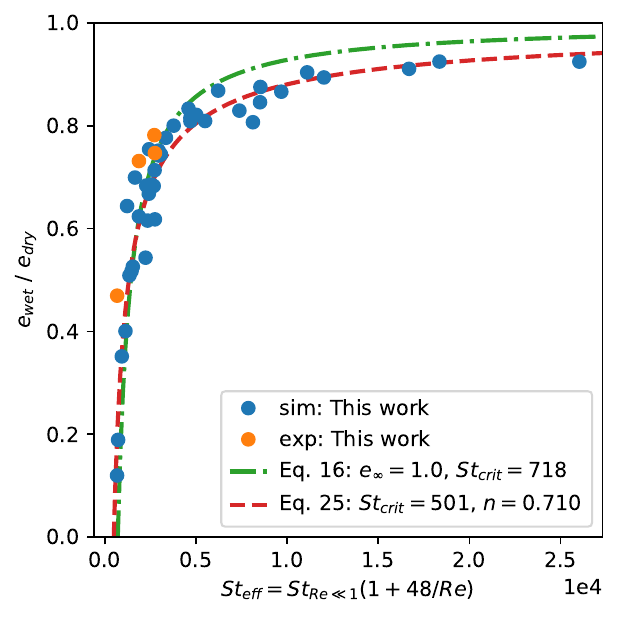}
   \caption{Linear scale}\label{fig::results::FittingStokes1::normal}
    \end{subfigure}  
    \begin{subfigure}[b]{0.48\textwidth}
   \centering
   \includegraphics[width=\textwidth]{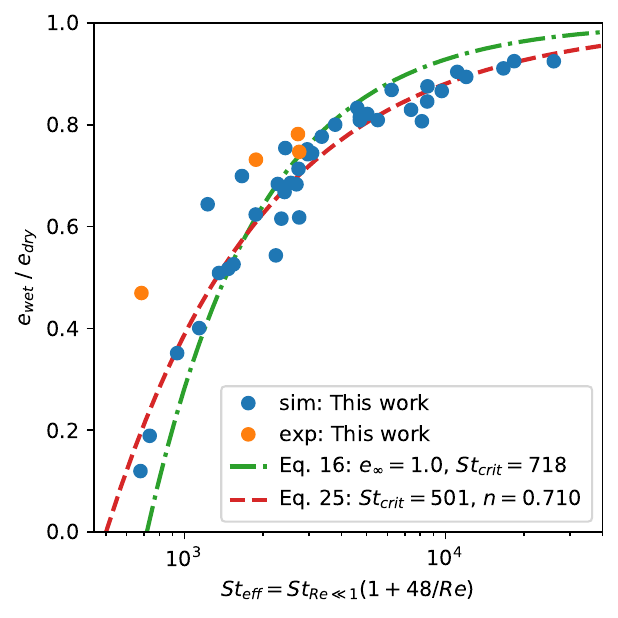}
   \caption{Logarithmic scale}\label{fig::results::FittingStokes1::logarithmic}
    \end{subfigure}  
    \caption{Fitting wet restitution coefficient from Eq. \ref{eq::StokesModelDavis} and Eq. \ref{eq::StokesModelDampened}, with $e_{\infty}$ fixed on 1.0, on simulation data, compared to simulations and experiments}\label{fig::results::FittingStokes1}
\end{figure}

Based on the new definition of the Stokes number in equation \ref{eq::resutls::newStokesNumberViscInertia}, all data can be described by a single predictive closure correlation. 
Note that previous works on defining such closure relations for these wet-particle collisions focused on two major routes: liquid-bridge force modeling and correlating the restitution coefficients. 
The most common liquid-bridge force models are the Easo \cite{Easo2013LiquidBridge} and Washino \cite{WASHINO2016100} model. 
These models are derived from an analytical description of the liquid bridge in the pendular regime (liquid content < 10\%), while the current simulations have significantly higher liquid contents. 
This difference is visible when comparing these models with the default values from LIGGGHTS \cite{liggghts_easo_capillary_viscous} with the simulations from this work.
As these models describe only viscous forces (which depend on viscosity and slip velocity) and capillary forces (which are negligible in this regime), the models are unable to describe the current simulation set (RMSE = 17.2\% and 28.9\%, respectively). The interested reader is referred to
\ref{sec::appendix0::washinoisbad} for more details. This was already established by \citet{STROHNER2026387} when $Re>1$.\\
Consequently, we propose to describe the collision dynamics with an effective restitution coefficient. 
Figure \ref{fig::results::FittingStokes1} shows Eq. \ref{eq::StokesModelDavis} fitted to the data of the parametric study and shows the head-on experiments of this work. 
The new Stokes number definition collapses of these results on a single trend independent of the exact physical properties and the original distribution of the liquid. Additionally it collapses the experimental validation results.
Although the agreement between the model in Equation \ref{eq::StokesModelDavis} and the simulations is reasonable, further improvements are possible. $e_{\infty}$ was fixed at 1.0 to satisfy the high-Stokes-numbers limit, as the restitution coefficient was normalized by the dry restitution coefficient. Allowing $e_{\infty}$ to decrease below unity would reduce the RMSE from 7.39\% to 5.03\%. However, this would violate the intrinsic normalization of the restitution coefficient.
To improve the fit while preserving these constraints, a damping exponent is introduced in Equation \ref{eq::StokesModelDampened}.
\begin{equation}\label{eq::StokesModelDampened}
    e_{wet}=e_{\infty}\left(1-\left(\frac{St_{crit}}{St_{eff}}\right)^n  \right)
\end{equation}
In this formulation, the $St_{crit}$ still describes the onset of agglomeration and the $e_{\infty}$ still describes the asymptote the graph approaches at high St, being unity.
$n$ is the exponent that controls the growth rate of the function. Please note the relation will employ the $St_{eff}$ from Equation \ref{eq::resutls::newStokesNumberViscInertia}.
Figure \ref{fig::results::FittingStokes1} also shows Equation \ref{eq::StokesModelDampened}, where $e_{\infty}$ is fixed at 1.0. The critical Stokes number is 501, which is in similar range as for Equation \ref{eq::StokesModelDavis} with 718.
The exponent $n$ is 0.714, showing its importance in controlling the shape of the fit. 
The physical interpretation of $n$ remains the subject of future investigation. However, this new model can be very valuable in describing the restitution coefficient for wet particle collisions. This modified model reduces the RMSE from 7.67\% to 4.99\%.

\section{Conclusion}
In this study, a Direct Numerical Simulations framework for wet particle collisions was validated against newly performed experiments for two wet particles and demonstrates good agreement in both liquid-bridge evolution and rebound dynamics. The earlier rupture of the liquid bridge does not significantly influence the rebound of particle. The framework was subsequently employed to systematically investigate the effects of liquid viscosity, liquid density, liquid-layer thickness, impact velocity, and surface tension coefficient on the wet restitution coefficient.

The results show that increasing liquid viscosity, liquid content, and liquid-to-solid density ratio reduces the restitution coefficient, whereas increasing impact velocity promotes rebound. The effect of surface tension is negligible in these ranges of conditions, indicating that viscous dissipation is the primary mechanism in reducing the kinetic energy.

To unify the observed trends, existing definitions of the Stokes number were evaluated. Although these definitions capture individual parameter dependencies, none were able to simultaneously include all the aforementioned effects. A new generalized Stokes number introduces a new characteristic velocity based on the flow in the gap between the particles and combines the characteristic time scales associated with response time at creeping-flow ($Re\ll1$) and inertia-dominated ($Re>1000$) conditions. This definition successfully collapses the data to a common trend and is used to fit an empirical correlation (with a RMSE of 4.99\%) with three parameters: A critical Stokes number associated with the onset of agglomeration, an exponent that governs the shape of the correlation, and the asymptotic restitution coefficient at high Stokes numbers, which is put to the dry restitution coefficient. This model can be implemented in CFD-DEM simulations as a closure relation for the effect the liquid layer for wet particle collisions in fluidized beds.

\section{Acknowledgment}
This publication is part of the project Understanding of wet agglomeration of non-spherical particles in fluidized beds with project number 19951 of the Open Technology Programme, which is financed by the Dutch Research Council (NWO). This work was funded by the Deutsche Forschungsgemeinschaft (DFG, German Research Foundation) – project number 438775980. This work used the Dutch national e-infrastructure with the support of the SURF Cooperative using grant no. EINF-15093 and EINF-19266.
\section{Data availability}
The data underlying this publication and the associated data manipulation scripts are published in 
\url{https://doi.org/10.4121/951cce83-73ca-4236-9756-d9cc54208adf}.

\section{AI statement}
AI tools (GPT-5.4, Opus-5.5) were used to assist with generating plotting code and for spelling and grammar checks during manuscript preparation. AI was not used for data analysis, interpretation of results, or scientific writing. The authors reviewed and verified all outputs and take full responsibility for the content of the manuscript.
\bibliographystyle{plainnat}
\bibliography{references}
\appendix
\section{Comparison with liquid bridge force models}\label{sec::appendix0::washinoisbad}
\begin{figure}
    \centering
    \includegraphics[width=0.5\linewidth]{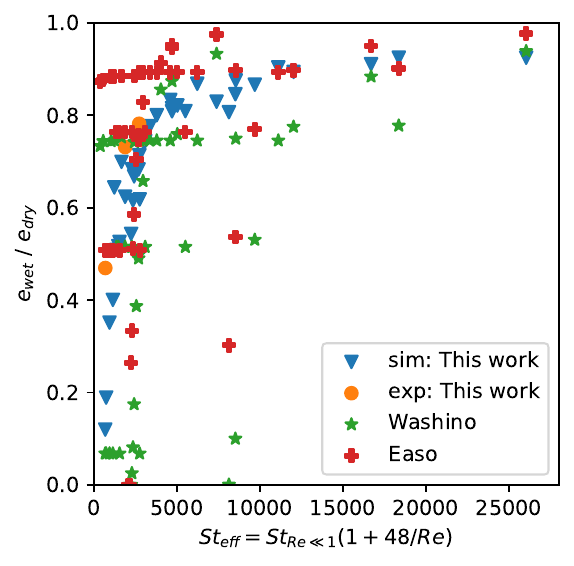}
    \caption{Comparison Washino \cite{WASHINO2016100} and Easo \cite{Easo2013LiquidBridge} model with the simulations and experiments from this work.}
    \label{fig:appendix1:washino}
\end{figure}

In section \ref{sec::results::stokesmodel}, the performed simulations in this work are compared with the results from Easo and Washino model \cite{WASHINO2016100,Easo2013LiquidBridge}. The default settings for the Easo and Washino model from LIGGGHTS have been used \cite{liggghts_easo_capillary_viscous}. Figure \ref{fig:appendix1:washino} shows this comparison. The horizontal alignment of the results from the Easo and  Washino model clearly indicate these models fail to include the effect of liquid content, liquid density, and solid density. Additionally, the predicted onset of agglomeration, i.e. the $St_{crit}$, by these models is significantly different from the results presented in this work.  
The RMSE of 17.2\% and 28.9\% for the Easo and Washino model, respectively, indicate that these models are not applicable in this range.

\section{Restitution coefficient data}\label{sec::appendix1::rawdata}
The restitution coefficient for various conditions, from the result section is shown in Table \ref{appendix1:rawdata_experiments} fro the experiments and Table \ref{appendix1:rawdata} for the simulations.

\begin{table}[H]
    \centering
    \caption{Restitution coefficient for various conditions from experiments.}\label{appendix1:rawdata_experiments}
    \resizebox{0.9\textwidth}{!}{%
\begin{tabular}{cccccccccc}
 \hline
$\gamma$ [$^\circ$] & $V_{l,1}/V_{s,1}$ [\%] & $V_{l,2}/V_{s,2}$ [\%] & $U_p$ [m/s] & $\rho_l$ [kg/m$^3$] & $\rho_p$ [kg/m$^3$] & $\mu_l$ [Pa$\cdot$s] & $\sigma$ [N/m] & $St_{new}$ & $e$ \\ \hline
175.0 & 60.5 & 33.4 & 1.79 & 968 & 6050 & 0.1 & 0.021 & 1879 & 0.731 \\
166.4 & 87.0 & 51.4 & 1.79 & 951 & 6050 & 0.02 & 0.0201 & 2754 & 0.747 \\
165.3 & 80.8 & 57.6 & 1.78 & 951 & 6050 & 0.02 & 0.0201 & 2723 & 0.782 \\
168.3 & 168.1 & 77.6 & 1.51 & 960 & 6050 & 0.05 & 0.0201 & 684 & 0.470 \\
\hline
 \end{tabular}
    }
\end{table}

\begin{table}[H]
    \centering
    \caption{Restitution coefficient for various conditions from simulations.}\label{appendix1:rawdata}
    \resizebox{0.9\textwidth}{!}{%
\begin{tabular}{cccccccccc}
 \hline
$V_{l,1}/V_{s,1}$ [\%] & $V_{l,2}/V_{s,2}$ [\%] & $U_p$ [m/s] & $\rho_l$ [kg/m$^3$] & $\rho_p$ [kg/m$^3$] & $\mu_l$ [Pa$\cdot$s] & $\sigma$ [N/m] & $St_{new}$ & $e$ \\ \hline
33.1 & 33.1 & 2.00 & 968 & 6050 & 0.1 & 0.021 & 3359 & 0.777 \\
15.8 & 15.8 & 2.00 & 968 & 6050 & 0.1 & 0.021 & 12019 & 0.894 \\
52.1 & 52.1 & 2.00 & 968 & 6050 & 0.1 & 0.021 & 1661 & 0.699 \\
33.1 & 33.1 & 2.00 & 968 & 6050 & 0.025 & 0.021 & 7385 & 0.829 \\
33.1 & 33.1 & 2.00 & 968 & 6050 & 0.05 & 0.021 & 4701 & 0.809 \\
33.1 & 33.1 & 2.00 & 968 & 6050 & 0.2 & 0.021 & 2688 & 0.683 \\
33.1 & 33.1 & 2.00 & 968 & 6050 & 0.4 & 0.021 & 2352 & 0.616 \\
33.1 & 33.1 & 2.00 & 968 & 6050 & 0.6 & 0.021 & 2240 & 0.543 \\
15.8 & 15.8 & 2.00 & 968 & 6050 & 0.025 & 0.021 & 26026 & 0.925 \\
15.8 & 15.8 & 2.00 & 968 & 6050 & 0.05 & 0.021 & 16688 & 0.911 \\
15.8 & 15.8 & 2.00 & 968 & 6050 & 0.2 & 0.021 & 9685 & 0.866 \\
15.8 & 15.8 & 2.00 & 968 & 6050 & 0.4 & 0.021 & 8517 & 0.846 \\
15.8 & 15.8 & 2.00 & 968 & 6050 & 0.6 & 0.021 & 8128 & 0.807 \\
33.1 & 33.1 & 2.00 & 200 & 6050 & 0.1 & 0.021 & 11103 & 0.904 \\
33.1 & 33.1 & 2.00 & 400 & 6050 & 0.1 & 0.021 & 6222 & 0.868 \\
33.1 & 33.1 & 2.00 & 600 & 6050 & 0.1 & 0.021 & 4596 & 0.833 \\
33.1 & 33.1 & 2.00 & 800 & 6050 & 0.1 & 0.021 & 3782 & 0.800 \\
33.1 & 33.1 & 2.00 & 1200 & 6050 & 0.1 & 0.021 & 2969 & 0.742 \\
33.1 & 33.1 & 2.00 & 1400 & 6050 & 0.1 & 0.021 & 2736 & 0.713 \\
33.1 & 33.1 & 2.00 & 200 & 3000 & 0.1 & 0.021 & 5505 & 0.809 \\
33.1 & 33.1 & 2.00 & 400 & 3000 & 0.1 & 0.021 & 3085 & 0.744 \\
33.1 & 33.1 & 2.00 & 600 & 3000 & 0.1 & 0.021 & 2279 & 0.684 \\
33.1 & 33.1 & 2.00 & 800 & 3000 & 0.1 & 0.021 & 1875 & 0.624 \\
33.1 & 33.1 & 2.00 & 1200 & 3000 & 0.1 & 0.021 & 1472 & 0.517 \\
33.1 & 33.1 & 2.00 & 1400 & 3000 & 0.1 & 0.021 & 1357 & 0.509 \\
33.1 & 33.1 & 2.00 & 200 & 1500 & 0.1 & 0.021 & 2753 & 0.618 \\
33.1 & 33.1 & 2.00 & 400 & 1500 & 0.1 & 0.021 & 1543 & 0.526 \\
33.1 & 33.1 & 2.00 & 600 & 1500 & 0.1 & 0.021 & 1139 & 0.401 \\
33.1 & 33.1 & 2.00 & 800 & 1500 & 0.1 & 0.021 & 938 & 0.351 \\
33.1 & 33.1 & 2.00 & 1200 & 1500 & 0.1 & 0.021 & 736 & 0.189 \\
33.1 & 33.1 & 2.00 & 1400 & 1500 & 0.1 & 0.021 & 678 & 0.120 \\
33.1 & 33.1 & 2.00 & 968 & 6050 & 0.1 & 0.005 & 3359 & 0.780 \\
33.1 & 33.1 & 2.00 & 968 & 6050 & 0.1 & 0.01 & 3359 & 0.791 \\
33.1 & 33.1 & 2.00 & 968 & 6050 & 0.1 & 0.04 & 3359 & 0.777 \\
33.1 & 33.1 & 0.60 & 968 & 6050 & 0.1 & 0.021 & 2419 & 0.668 \\
33.1 & 33.1 & 0.80 & 968 & 6050 & 0.1 & 0.021 & 2554 & 0.686 \\
33.1 & 33.1 & 1.40 & 968 & 6050 & 0.1 & 0.021 & 2956 & 0.752 \\
33.1 & 33.1 & 4.00 & 968 & 6050 & 0.1 & 0.021 & 4701 & 0.815 \\
33.1 & 15.8 & 0.40 & 968 & 6050 & 0.1 & 0.021 & 3869 & 0.733 \\
33.1 & 15.8 & 0.60 & 968 & 6050 & 0.1 & 0.021 & 4092 & 0.731 \\
33.1 & 15.8 & 0.80 & 968 & 6050 & 0.1 & 0.021 & 4315 & 0.791 \\
33.1 & 15.8 & 1.00 & 968 & 6050 & 0.1 & 0.021 & 4538 & 0.811 \\
33.1 & 15.8 & 1.40 & 968 & 6050 & 0.1 & 0.021 & 4983 & 0.833 \\
33.1 & 15.8 & 3.00 & 968 & 6050 & 0.1 & 0.021 & 6764 & 0.862 \\
33.1 & 15.8 & 4.00 & 968 & 6050 & 0.1 & 0.021 & 7878 & 0.871 \\
33.1 & 0.0 & 0.40 & 968 & 6050 & 0.1 & 0.021 & 8284 & 0.736 \\
33.1 & 0.0 & 0.60 & 968 & 6050 & 0.1 & 0.021 & 8751 & 0.750 \\
33.1 & 0.0 & 0.80 & 968 & 6050 & 0.1 & 0.021 & 9218 & 0.810 \\
33.1 & 0.0 & 1.00 & 968 & 6050 & 0.1 & 0.021 & 9685 & 0.834 \\
33.1 & 0.0 & 1.40 & 968 & 6050 & 0.1 & 0.021 & 10618 & 0.845 \\
33.1 & 0.0 & 3.00 & 968 & 6050 & 0.1 & 0.021 & 14354 & 0.889 \\
33.1 & 0.0 & 4.00 & 968 & 6050 & 0.1 & 0.021 & 16688 & 0.888 \\
12.5 & 12.5 & 2.00 & 968 & 6050 & 0.1 & 0.021 & 18356 & 0.925 \\
19.1 & 19.1 & 2.00 & 968 & 6050 & 0.1 & 0.021 & 8538 & 0.875 \\
26.0 & 26.0 & 2.00 & 968 & 6050 & 0.1 & 0.021 & 5022 & 0.821 \\
40.5 & 40.5 & 2.00 & 968 & 6050 & 0.1 & 0.021 & 2436 & 0.754 \\
64.3 & 64.3 & 2.00 & 968 & 6050 & 0.1 & 0.021 & 1228 & 0.644 \\
15.8 & 0.0 & 2.00 & 968 & 6050 & 0.1 & 0.021 & 45391 & 0.968 \\
33.1 & 0.0 & 2.00 & 968 & 6050 & 0.1 & 0.021 & 12019 & 0.942 \\
52.1 & 0.0 & 2.00 & 968 & 6050 & 0.1 & 0.021 & 5651 & 0.825 \\
72.8 & 0.0 & 2.00 & 968 & 6050 & 0.1 & 0.021 & 3359 & 0.788 \\
119.7 & 0.0 & 2.00 & 968 & 6050 & 0.1 & 0.021 & 1661 & 0.692 \\
174.4 & 0.0 & 2.00 & 968 & 6050 & 0.1 & 0.021 & 1036 & 0.674 \\
15.8 & 7.7 & 2.00 & 968 & 6050 & 0.1 & 0.021 & 20765 & 0.931 \\
33.1 & 15.8 & 2.00 & 968 & 6050 & 0.1 & 0.021 & 5651 & 0.847 \\
52.1 & 24.2 & 2.00 & 968 & 6050 & 0.1 & 0.021 & 2727 & 0.754 \\
72.8 & 33.1 & 2.00 & 968 & 6050 & 0.1 & 0.021 & 1661 & 0.683 \\
\hline
 \end{tabular}
    }
\end{table}
\section{Supplementary videos} 
\begin{figure}[h]
    \centering
    \textcolor{red}{Include case01\_4s.mp4 here}
    \includegraphics[width=0.95\linewidth]{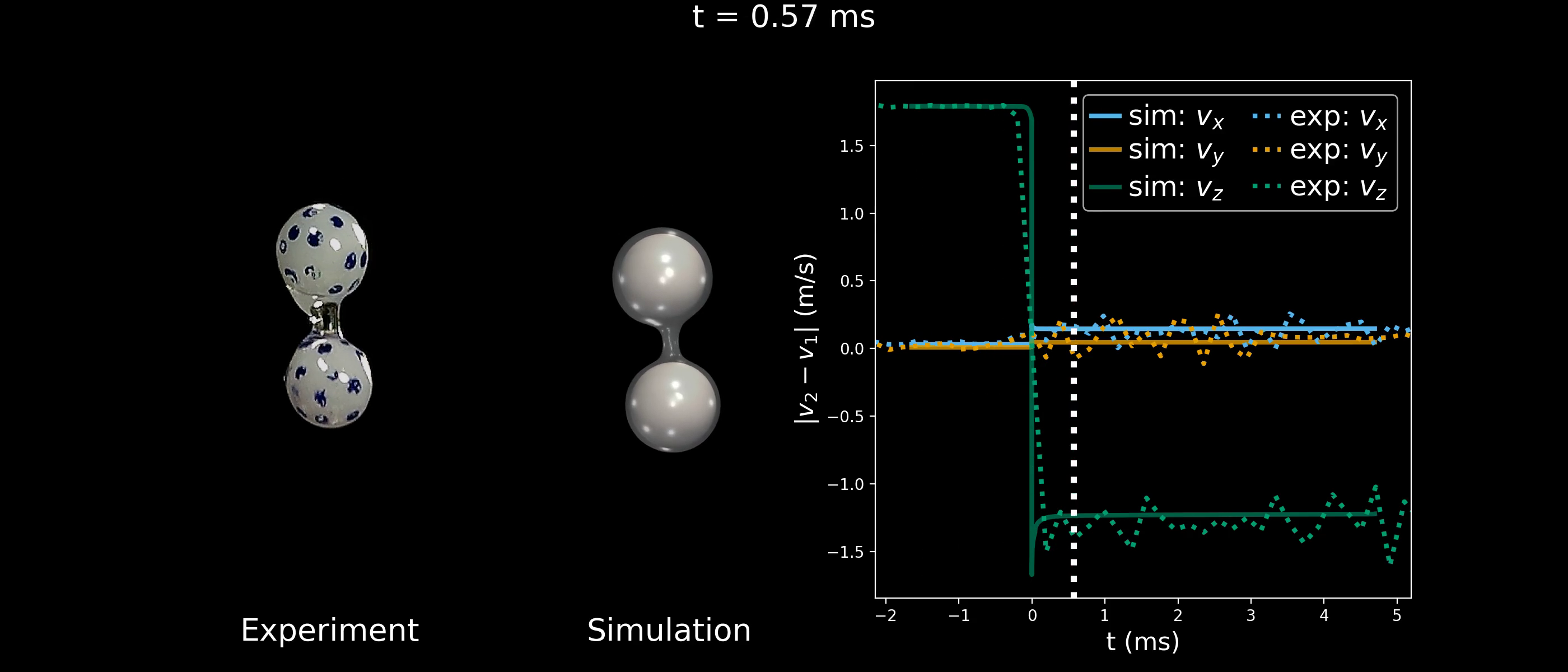}
    \caption{Video of collision at collision angle 175.0\textdegree for both experiments (Exp) and simulations (Sim) along with the comparison of the translational velocity}
    \label{vid:case01}
\end{figure}

\begin{figure}[h]
    \centering
    \textcolor{red}{Include case30\_4s.mp4 here}
    \includegraphics[width=.95\linewidth]{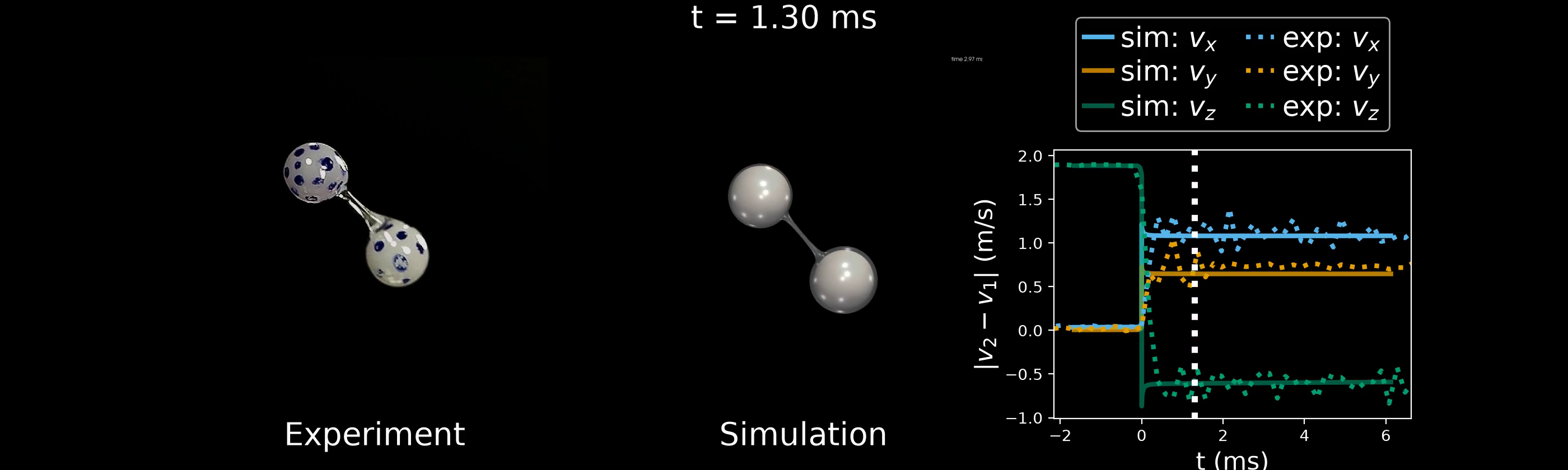}
    \caption{Video of collision at collision angle 146.5\textdegree for both experiments (Exp) and simulations (Sim) along with the comparison of the translational velocity}
    \label{vid:case30}
\end{figure}

\begin{figure}[h]
    \centering
    \textcolor{red}{Include case44\_4s.mp4 here}
    \includegraphics[width=0.95\linewidth]{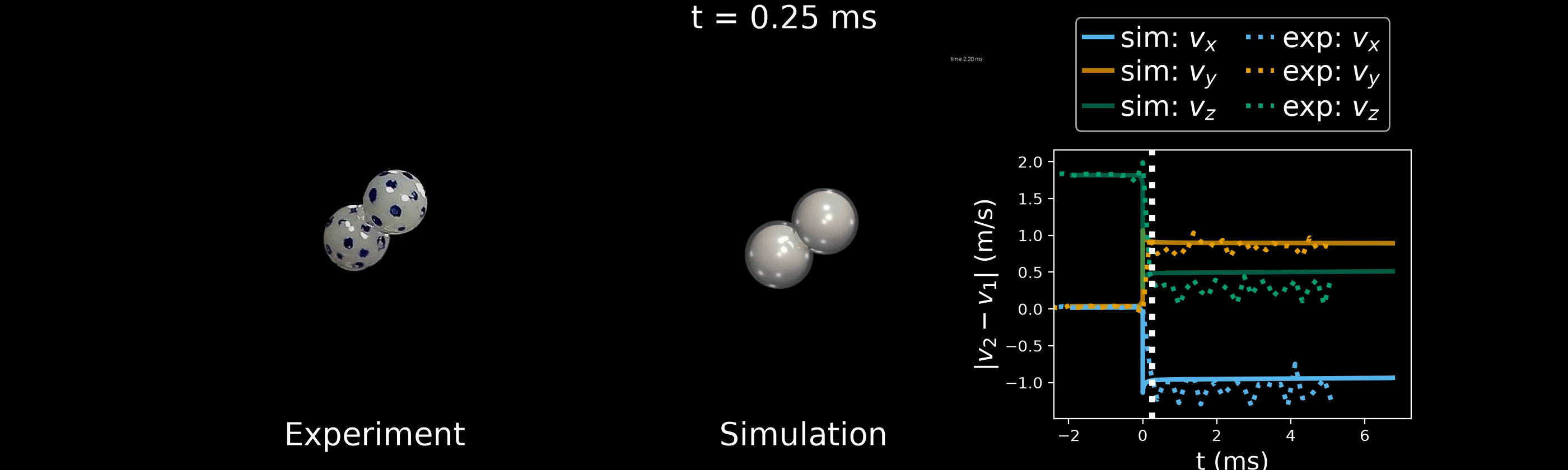}
    \caption{Video of collision at collision angle 136.1\textdegree for both experiments (Exp) and simulations (Sim) along with the comparison of the translational velocity}
    \label{vid:case44}
\end{figure}

\textcolor{red}{*** Only online available ***} \\ \\
The videos of the collisions shown in Figures \ref{fig::validation::case01::sidebyside} and \ref{fig::validation::case01::velocities} are provided in this section for the interested reader. Video \ref{vid:case01} shows the collision at $\gamma = 175.0 \degree$, Video \ref{vid:case30} shows the collision at $\gamma = 146.5 \degree$, and Video \ref{vid:case44} shows the collision at $\gamma = 136.1 \degree$

\end{document}